\documentclass[twocolumn,showpacs,preprintnumbers,pre]{revtex4}
\usepackage{epsfig,amsmath,amssymb}

\begin{document}

\title{Phase-shifts in stochastic resonance in a Chua circuit}
\author{Wojciech Korneta}
\thanks{wkorneta@op.pl}
\affiliation{Faculty of Physics, Technical University, Malczewskiego 29, 26-600 Radom, Poland}
\author{Iacyel Gomes, Claudio R. Mirasso, Ra\'ul Toral}
\affiliation{IFISC (Instituto de f{\'\i}sica interdisciplinar y sistemas complejos), CSIC-Universitat de les Illes Balears, Ed. Mateu Orfila, Campus UIB, E-07122 Palma de Mallorca, Spain}
\date{\today}

\begin{abstract}
We present an experimental study of stochastic resonance in an electronic Chua circuit operating in the chaotic regime. We study in detail the switch-phase distribution and the phase-shift between sinusoidal forcing for two responses of the circuit: one depending on both inter-well and intra-well dynamics and the other depending only on inter-well dynamics. We describe the two relevant de-synchronizatrion mechanisms for high and low frequencies of the forcing and present a method to detect the optimal noise intensity from switch phases which coincides with the one derived from the observation of the signal-to-noise ratio or residence times.
\end{abstract}

\pacs{05.40.-a,02.50.Ey}
\maketitle
\section{Introduction}
Stochastic resonance, by which a weak signal acting upon a nonlinear system can be amplified by the addition of noise of optimal intensity, has attracted considerable attention in recent years. While most of the research has focused on bistable and excitable systems \cite{1,2}, studies have also been performed in chaotic systems with two attractors, in what can be considered as a generalized form of a bistable system \cite{3,4}. Several measures have been introduced to quantify the strength of the response to the forcing: the response amplitude originally used by Benzi et al.\cite{5}, the signal-to-noise ratio \cite{1,2}, as well as other measures determined from the residence and switching times probability distributions \cite{6,7,8} or the work done by the external force \cite{9}. Although all these different quantifiers display a maximum as a function of the noise intensity, the values of the optimal noise intensity obtained in each case do not necessarily coincide.

In both theoretical and experimental studies the phase-shift between the force and the response has been much less considered, although it is interesting from the viewpoint of relating stochastic resonance to standard resonance phenomena. Moreover, the magnitude of the phase-shift and its variation with noise intensity has been the subject of some controversy \cite{10,11,12}. The phase-shift has been analytically calculated in bistable system within the linear response theory taking into account both intra-well and inter-well dynamics \cite{10,13}. It displays a bell-shape dependence on the noise intensity with a extremum at a smaller value of the noise intensity than the optimal noise intensity obtained from the signal-to-noise ratio. This result was confirmed by measurements on an analog electronic circuit \cite{14} and the same dependence has also been obtained numerically \cite{9,15}. However, results in a one-dimensional Ising model show a monotonous change of the phase-shift with the noise intensity\cite{16}. This controversy \cite{11,12} was considered in Refs.\cite{16,17,18} based on numerical studies of a bistable system and an assembly of superparamagnetic particles. One obtains the monotonous change of the phase-shift with the noise intensity if only inter-well dynamics is taken into account. The bell-shape dependence thus reflects the competition between inter-well and intra-well dynamics. It was then concluded that the extrema in the dependence of the phase-shift and signal-to-noise ratio have a different origin. The phase-shift has also been studied in terms of switch-phase distributions in Ref.\cite{19}. The authors presented a numerical investigation in the symmetric Schmitt trigger and proposed the de-synchronization mechanism responsible for the disappearance of stochastic resonance.

The aim of this paper is to present experimental results of our thorough studies of the phase-shift in stochastic resonance in the Chua electronic circuit \cite{20} operating in a chaotic regime where two single-scroll attractors coexist. In this case the circuit can be thought of as bistable and the observed phenomena correspond to conventional stochastic resonance within two stable wells\cite{14}. We have determined the phase-shift between a sinusoidal forcing $E(t)=E_0\sin(\omega t)$ and two responses of the circuit: one depending on both inter-well and intra-well dynamics and the other depending only on inter-well dynamics. This clarifies the dependence of the phase-shift on the noise intensity and parameters of external forcing. Considering switching-phase distributions one can distinguish two de-synchronization mechanisms for low and high frequencies of external forcing. This leads us to propose different methods to determine the optimal noise intensity. We show that the same optimal noise intensity can be obtained from the signal-to-noise ratio and the switching, residence time or switching-phase distributions. Stochastic resonance is a widespread phenomenon and we hope that our methods will provide an easy and appropriate way to determine numerically or experimentally the optimal noise intensity.

\section{Experimental setup and results}

\begin{figure} \center
\epsfig{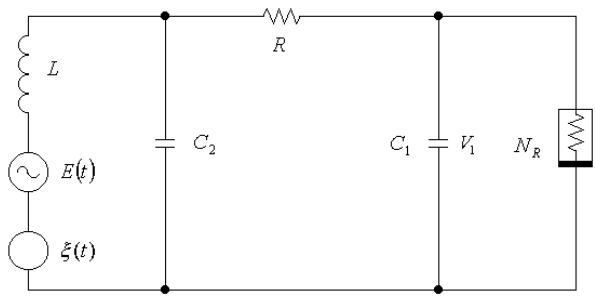}
\caption{\label{fig:1} Diagram of electronic Chua's circuit. $E(t)$ is the forcing sinusoidal signal and $\xi(t)$ is a Gaussian white noise source. $R=1675\Omega$, $L=18$mH , $C_1=10$nF and $C_2=100$nF . The nonlinear Chua's diode $N_R$ has the current-voltage characteristic: $I=f(V)=m_1V+\frac{1}{2}(m_0-m_1)(|V+V_0|-|V-V_0|)$, with parameters $m_0=-0.758$mA/V, $m_1=-0.409$mA/V and $V_0=1.08$V. $V_1$ is the voltage on the capacitor $C_1$ .}
\end{figure}

The Chua circuit and parameters of its components used in our experiments are given in Fig.\ref{fig:1}. Stochastic resonance in this circuit was observed and described in Ref.\cite{4}. In the absence of the forcing, $E(t)$, and the noise source, $\xi(t)$, the dynamics displays two single-scroll attractors and trajectories flow to one or another depending on initial conditions. Jumps between the attractors can be observed if the amplitude $E_0$ of the forcing is above a threshold value that depends on the frequency $\omega$ of the forcing \cite{4}. In all experiments we first set the amplitude of forcing just below that threshold value and then add the noise signal to the forcing so to induce jumps of the system between single-scroll attractors. During the experiments we recorded both $E(t)$ and the voltage $V_1(t)$ on the capacitor $C_1$. This voltage is selected because it clearly shows the position of the dynamical trajectory on either attractor and it represents the response of the circuit depending on both inter-well and intra-well dynamics. The step function representation, $S(t)$, of this voltage has been used in many studies \cite{1} as it represents the response of the circuit depending only on inter-well dynamics. 

\begin{figure} \center
\epsfig{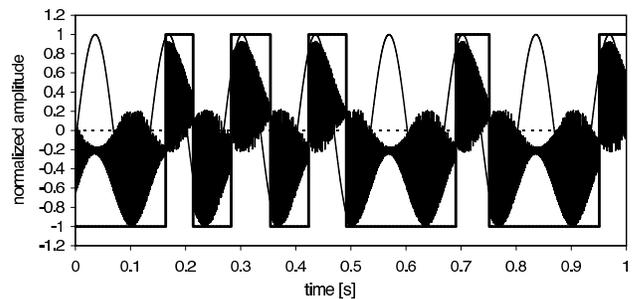}
\caption{\label{fig:2} Temporal evolution of the forcing $E(t)=E_0\sin(\omega t)$ and the voltage $V_1(t)$ with its step function representation $S(t)$. The amplitude of the forcing signal is $E_0=535$ mV and the frequency $\omega=7.5$ Hz.}
\end{figure}

\begin{figure} \center
\epsfig{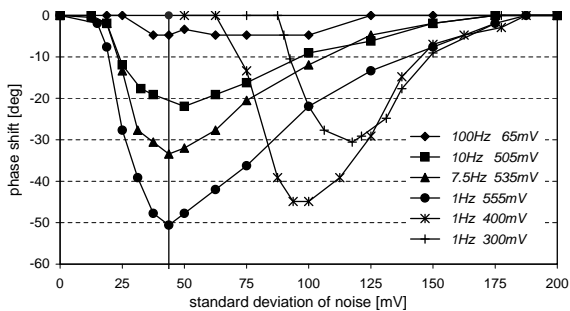}
\vspace{-9.0truecm}
\caption{\label{fig:3} The dependence on the noise intensity of the phase-shift $\Phi_{V_1}$ between the forcing $E(t)$ and the output voltage $V_1(t)$ for different amplitudes and frequencies.}
\end{figure}

We have computed phase-shifts both for $V_1(t)$ and $S(t)$. An example of temporal evolution of $E(t)$, $V_1(t)$ and $S(t)$ is shown in Fig.\ref{fig:2}. At forcing frequencies below $100$Hz one can determine the phase-shift between the forcing $E(t)$ and the response, e.g. $V_1(t)$, as the value $\Phi_{V_1}$ of the phase $\phi$ such that the cross-correlation function $\langle E(t)V_1(t+\phi/\omega)\rangle$ takes the maximum value (where $\langle\cdot\cdot\cdot\rangle$ represents a time average). In Fig. \ref{fig:3} we present the dependence of the phase-shift $\Phi_{V_1}$ on the noise intensity (measured as the standard deviation of its probabiity distribution) for different amplitudes $E_0$ and frequencies $\omega$. This is the bell-shape dependence observed and explained by Dykman et al. \cite{13}. For small noise levels, inter-well hoppings can be neglected and the circuit dynamics depends almost totally on the interaction between $E(t)$ and the intra-well motion. The characteristic frequency of the dynamics of our circuit at the single-scroll attractor is around $2740$Hz, so the phase-shift remains small. The abrupt decrease in the phase is associated with the onset of inter-well jumps. 

\begin{figure} \center
\epsfig{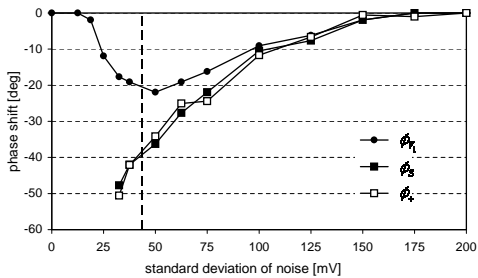}
\vspace{-10.0truecm}
\caption{\label{fig:4} Phase-shifts $\Phi_{V_1}$, $\Phi_S$ (defined from the correlation of the forcing $E(t)$ and the responses $V_1(t)$ and $S(t)$, respectively) and the average switch-phase $\Phi_+$ plotted as a function of noise intensity. The amplitude of the forcing is $E_0=505$mV and the frequency $\omega=10$Hz.}
\end{figure}

One can notice that the minimum of the phase-shift depends both on $E_0$ and $\omega$. This is illustrated by the data presented in Fig.\ref{fig:3} in the case $\omega=1$Hz. The threshold amplitude value in this case is about $555$mV. For lower forcing amplitudes higher noise intensities are necessary to induce jumps between attractors and the minimum of the bell-shape curve moves to higher noise intensities. At low $\omega$ the phase-shift $\Phi_S$ between $E(t)$ and the step function representation of the voltage $S(t)$ can also be determined from the maximum of the corresponding correlation function. An example of the dependence of this phase-shift on the noise intensity is shown in Fig.\ref{fig:4}  in the case $\omega=10$Hz. This dependence is monotonic and similar to what has been obtained in a one-dimensional Ising model \cite{16}. By superimposing in this figure also the phase-shift $\Phi_{V_1}$ one can notice that as soon as inter-well jumps are activated they immediately become dominant and cause the decrease of the phase-shift between $E(t)$ and $V_1(t)$. The inter-well hoppings also make the phase-shift tend to zero for higher noise intensities.

\begin{figure} \center
\epsfig{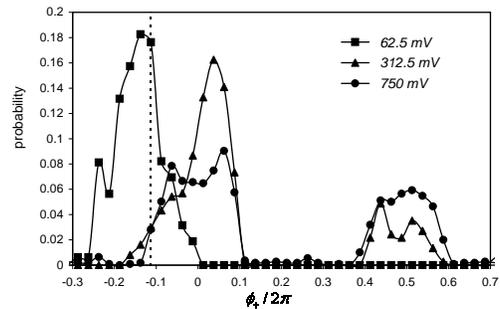}
\vspace{-8.0truecm}
\caption{\label{fig:5} Switch-phase distribution for different noise intensities (see the legend). The amplitude of the forcing signal is $535$mV and the frequency $7.5$ Hz. }
\end{figure}

Phase-shifts can also be determined from the switch-phase distribution. This method has been used in the symmetric Schmitt trigger \cite{19} and it is specially important for theoretical and experimental studies of stochastic resonance in neurons \cite{21}. In our case we looked at the distribution $f_+(\phi_+)$ of the oriented switch phases $\phi_+$ defined as the phases of the sinusoidal forcing signal (modulo $2\pi$) corresponding to switches of the step function $S(t)$ from negative to positive value. The distribution $f_-(\phi_-)$ of the switch phases $\phi_-$ in the opposite direction can be obtained by a translation by $\pi$ of the distribution of the switch phases $f_+(\phi_+)$. The average of the switch phases $\Phi_+$ is defined by the following equation:
\begin{equation}
\label{eq:1}
\rho \exp{(i\Phi_+)}=\frac{1}{N}\sum_{k=1}^{N}\exp{(i\phi_+^k)}
\end{equation}
where $\phi_+^k$, with $k=1, \dots, N$, are the different phase switches observed during the time evolution and the amplitude $\rho$ of the complex quantity serves as an order parameter characterizing the degree of phase synchronization in neurons \cite{21}. The dependence of $\Phi_+$ on the noise intensity coincides with that of the phase-shift $\Phi_S$ between $E(t)$ and $S(t)$, as shown in Fig.\ref{fig:4}. Both measures can thus be equivalently used to determine the phase-shift of a system whose response depends only on inter-well dynamics. In Fig.\ref{fig:5} we present distributions of switch phases $f_+(\phi_+)$ obtained at forcing frequency $\omega=7.5$Hz and different noise intensities. At low noise intensities there is only one peak in the distribution. As the noise intensity increases, this peak shifts towards zero and flattens. For a standard deviation of noise around $300$mV a second peak appears at distance $\pi$ from the dominating peak. The onset of the second peak signals the onset of de-synchronization in stochastic resonance. As the noise intensity increases the second peak grows until both peaks become comparable. This de-synchronization mechanism was observed and described in the numerical studies of the symmetric Schmitt trigger \cite{19} and signals the degradation of the stochastic resonance with the noise intensity.

\begin{figure} \center
\epsfig{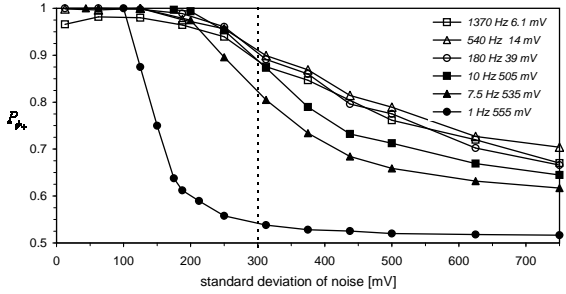}
\vspace{-9.0truecm}
\caption{\label{fig:6} The probability $P_{\Phi_+}$ that the switch-phase $\Phi_+$ is in the interval $[\Phi_+^D-\pi/2,\Phi_+^D+\pi/2]$ as a function of noise intensity. $\Phi_+^D$ denotes the position of the dominating peak in the switch-phase distribution. The legend gives the amplitudes and frequencies of the forcing periodic signal.}
\end{figure}

\begin{figure} \center
\epsfig{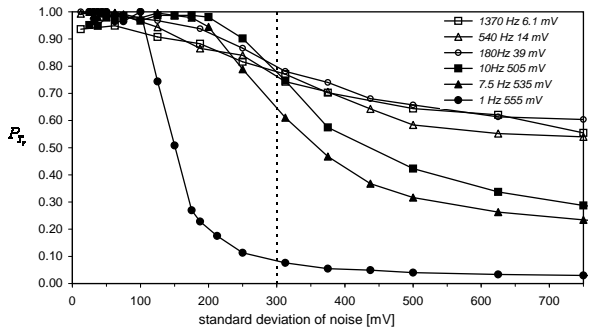}
\vspace{-9.0truecm}
\caption{\label{fig:7} The probability $P_{T_r}$ that the residence time is in one of intervals $[(i-3/4)\frac{2\pi}{\omega},(i-1/4)\frac{2\pi}{\omega}]$ , where $i=1,2,\dots$, as a function of noise intensity. The legend gives the amplitudes and frequencies of the forcing periodic signal}
\end{figure}

These observations allow us to propose a quantity which can be used to determine the optimal noise level from switch-phase distributions. Let us denote by $P_{\phi_+}$ the probability that the switch-phase $\phi_+$ belongs to the interval $[\phi_+^D-\pi/2,\phi_+^D+\pi/2]$, where $\phi_+^D$ is the position of the dominating peak in the switch-phase distribution. The dependence of this probability on the noise intensity for different forcing frequencies is presented in Fig.\ref{fig:6}. For frequencies higher than $7.5$Hz all curves in this figure have the inflection point at a noise intensity of $300$mV.  In a previous paper \cite{4} we considered the residence time distributions and defined an appropriate quantity which is suitable to detect the optimal noise level from these distributions. This quantity is the probability $P_{T_r}$ that the residence time is in one of intervals $[(i-3/4)\frac{2\pi}{\omega},(i-1/4)\frac{2\pi}{\omega}]$, where $i=1,2,\dots$. The dependence of this probability on the noise intensity for the same forcing frequencies as in Fig.\ref{fig:6} is shown in Fig.\ref{fig:7}. All curves at this figure also have the inflection point for the same noise level as curves in Fig.\ref{fig:6}. We observed \cite{4} that this noise level is the same as the optimal noise level obtained experimentally from the dependence of signal-to-noise ratio on noise intensity. We thus propose to use alternatively the dependence of the signal-to-noise ratio on the noise intensity or the dependence of probabilities $P_{\Phi_+}$ or $P_{T_r}$ on the noise intensity. In the first case the optimal noise level corresponds to the maximum whereas in the other cases it corresponds to the inflection point.

\begin{figure} \center
\epsfig{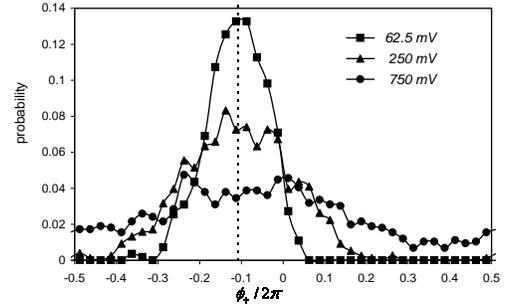}
\vspace{-9.0truecm}
\caption{\label{fig:8} Switch-phase distribution $f_+(\phi_+)$ for different noise intensities. The amplitude of the forcing signal is $14$mV and the frequency $540$Hz. The standard deviations of noise are given in the legend.}
\end{figure}

\begin{figure} \center
\epsfig{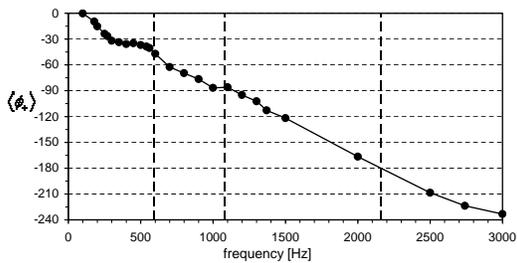}
\vspace{-9.0truecm}
\caption{\label{fig:9} The dependence of the average switch-phase $\Phi_+$ on the frequency of the forcing sinusoidal signal with an amplitude close to the threshold amplitude. }
\end{figure}

The methods described above to determine the phase-shift and de-synchronization mechanism change for forcing signals with frequencies above $100$Hz. In this case, the correlation function between the response of the system (either $V_1(t)$ or $S(t)$) and the forcing $E(t)$ vanishes and  the synchronization between the forcing and the response can only be determined from the switch-phase distributions, as the ones presented in Fig.\ref{fig:8} for different noise intensities and $\omega=540$Hz. One can note that there is only one peak in the distribution for all noise intensities. Increasing the noise intensity the position of the peak does not change but the distribution flattens. Finally for very large noise intensities the distributions of switch phases $\phi_+$ and $\phi_-$ start to overlap. This is the second mechanism of de-synchronization that ensues stochastic resonance on increasing further the noise level. The position of the peak in the switch-phase distribution changes with the forcing frequency. We present in Fig.\ref{fig:9} the dependence on the forcing frequency of the average switch-phase $\Phi_+$, as obtained from Eq. (\ref{eq:1}). The values $\Phi_+=-\pi/8,-\pi/4,-\pi/2$ correspond, respectively,  to frequencies $\omega=590$Hz, $1080$Hz, $2160$Hz. This dependence does not depend on the value of the sub-threshold amplitude of the forcing signal. It is evident from Figs. \ref{fig:5} and \ref{fig:6} that also for high frequencies the optimal noise intensity is determined by the inflection point of the dependence of the probability $P_{\phi_+}$ or $P_{T_r}$ on the noise intensity. As we have shown in Ref.\cite{4} this corresponds to the optimal noise intensity as determined by the maximum of the signal-to-noise ratio.

\section{Conclusions}

In this paper we have addressed the study of the the phase-shift in stochastic resonance. We have analyzed two mechanism of de-synchronization observed at high and low forcing frequencies. Our results point out the attention to switch-phases which are able quantify the synchronization between the forcing and response in the whole frequency range of the forcing signal. This paper also suggests methods to detect the optimal noise intensity by observing the signal-to-noise ratio, the residence times, the switching times or switch-phases. The optimal noise intensity determined from any of these observations is the same and they can be used alternatively. We have derived our main results by studying experimentally a Chua circuit operating in a chaotic regime, but we believe that our main conclusions are rather general and could be useful in many investigations and applications of the phenomenon of stochastic resonance.

\begin{acknowledgments}
We acknowledge financial support by the MEC (Spain) and FEDER (EU) through projects FIS2006-09966 and  FIS2007-60327.
\end{acknowledgments}

\end{document}